From: jwc@wuphys.wustl.edu
Sent: Tuesday, April 30, 2002 4:17 PM
To: psp
Subject: control.tex

\documentstyle[preprint,aps]{revtex}
\input epsf

\begin{document}
\draft
\title{Control of Quantum Systems}

\author{John W. Clark}

\address{Department of Physics,
Washington University, St. Louis, MO 63130 USA\\E-mail: jwc@wuphys.wustl.edu}
\author{Dennis G. Lucarelli and Tzyh-Jong Tarn}

\address{Department of Systems Science \& Mathematics,
Washington University,\\ 
St. Louis, MO 63130 USA\\E-mail: luc@zach.wustl.edu
\& tarn@wuauto.wustl.edu}

\maketitle
\begin{abstract}
A quantum system subject to external fields is said to be
controllable if these fields can be adjusted to guide the state
vector to a desired destination in the state space of the system.
Fundamental results on controllability are reviewed against
the background of recent ideas and advances in two seemingly
disparate endeavors: (i) laser control of chemical reactions and
(ii) quantum computation.  Using Lie-algebraic methods, 
sufficient conditions have been derived for global controllability on a
finite-dimensional manifold of an infinite-dimensional Hilbert
space, in the case that the Hamiltonian and control operators, possibly 
unbounded, possess a common dense domain of analytic vectors.  Some 
simple examples are presented.  A synergism between quantum 
control and quantum computation is creating a host of exciting new 
opportunities for both activities.  The impact of these developments 
on computational many-body theory could be profound.
\end{abstract}
\section{Introduction}

Since the birth of quantum theory, scientists have engaged in an 
abundance of experimental efforts to control quantum systems, 
with prominent successes in particle acceleration and detection, 
magnetic resonance, electron microscopy, solid-state electronics,
and laser trapping.  However, the need for a comprehensive theory
of quantum control was not recognized until the beginning of the
1980's, when powerful concepts and methods from systems engineering 
were first integrated into the quantum framework 
\cite{butkovskii,modeling,htc,ong1,ong2,inria}.  The subject 
soon received a strong impetus from the prospects for optical
control of chemical reactions opened by tremendous advances in
laser technology.  There followed a period of rapid development
of techniques for practical control of molecular dynamics, 
including continuous-wave coherent control \cite{brumer1}, 
dual-pulse control \cite{rice1}, optimal control \cite{rabitz1,rice2},
inverse control \cite{rabitz2}, and ``closed-loop'' learning
control \cite{rabitz3}.  This work has already been the subject of
several reviews \cite{rice3,brumer2,warren,clark,rabitz4}. 

Looking back to the origins of quantum control theory, it 
noteworthy that the seminal ideas of quantum information 
theory and quantum computation emerged during the same period 
\cite{benioff,feynman,deutsch1}.  The source of the
vast potential of both quantum control and quantum computation
stems from the superposition principle of quantum mechanics and
in turn from the simple fact that ``Hilbert space is a big place.''
The icon is the double-slit experiment, in which two quantum
paths are generated and caused to produce a wave interference
pattern.  More specifically:  
\begin{itemize}
\item[$\bullet$]
In {\sl quantum control} of molecular dynamics using lasers, one 
seeks to create two or more independent quantum pathways of the 
light-field--molecule system that interfere constructively 
or destructively so as to attain a specified target condition, 
e.g.~a larger yield for one reaction product in preference to others.    
``Manipulation of the phases of molecular, atomic, and electronic
systems, through the use of laser phase, provides a general paradigm
for control of quantum dynamics.'' \cite{brumersciam}
\item[$\bullet$]
Similarly, {\sl quantum computation} exploits superposition to
achieve massive parallelism and ideally an exponential speedup
of processing. ``A quantum computer obeys the laws of quantum mechanics, 
and its unique feature is that it can follow a superposition of [many!]
computational paths simultaneously and yield a final state
depending on the interference of these paths.'' \cite{cirac}.
\end{itemize}

Surprisingly, the intimate connection of these two developments has 
only recently been brought to the surface \cite{clark,lloyd1,rama1}.
Nevertheless it is clear that quantum control, in the sense of
implementing designated unitary transformations in the state space,
is an essential ingredient of quantum computation; and on the other hand,
quantum control theory will surely benefit from advances in quantum 
computing.

The mysteries and paradoxes of quantum mechanics, most especially
those arising from superposition and entanglement (the double-slit
experiment; Schr\"odinger's cat; the EPR paradox with its ``spooky
action at a distance'') are being harnessed to create the advanced
technologies of the future.  We are reminded of science-fiction
author Robert A. Heinlein's penetrating remark that ``Any advanced 
technology is indistinguishable from magic.''

\section{Controllability 101}

We begin our discussion of quantum dynamics and its
control at a textbook level, then increase or decrease
rigor as deemed convenient or appropriate.  The solution of the 
time-dependent Schr\"odinger equation 
\begin{equation}
i\hbar{\dot \psi(t)} = H \psi(t) 
\end{equation}
for the state vector $\psi(t)$ can be expressed in terms of a 
unitary transformation $\psi(t)= U\psi_0$ on the initial state 
$\psi (t=0) = \psi_0$.  Importantly, for a Hamiltonian $H$ that 
is not an explicit function of time, the unitary time-evolution 
operator $U$ takes the exponential form
\begin{equation}
U(t,0) = \exp \left[ -i H t/\hbar \right] \, .
\label{eq:uexp}
\end{equation}
Henceforth we take $\hbar =1$. 

To gain an intuitive grasp of the controllability problem, it is
instructive to consider a Schr\"odinger equation of the type
\begin{equation}
i {\dot \psi}(t) = H \psi(t) =
\sum_{i=1}^r u_i(t) H_i \psi(t)  \,, 
\label{eq:nodrift}
\end{equation}
again with boundary condition
$\psi(t=0)=\psi_0$, where the $u_i(t)$ are real control functions 
that can be turned on or off at will and the Hermitian control operators 
$H_i$ are all time independent.  Now let's go on a little trip, first 
here on Earth in a flat place like Holland and then in Hilbert space.  
Suppose in both cases there 
are only two controls, two possible ```directions'' of motion (four
when you count the possibilities of forward and backward
motion).  For the trip on Earth, go North for one hour, then
West for one hour, then South for one hour, and finally East
for one hour, maintaining a steady speed on each leg.  You wind up 
where you started, so the succession of translations has not given 
anything new (this being no surprise, since translation is a commutative 
operation).  

For the trip in Hilbert space, turn on $H_2$ for $t$ units of time, 
then $H_1$ for $t$ units, then $-H_2$ for $t$ units, and
finally $-H_1$ for $t$ units (the minus signs come from
the flexibility of the $u_i(t)$, which we take to be piecewise constant).  
In general, you do not wind up in the same place!   The state vector 
will reach the point
\begin{equation}
\psi = \exp(iH_1t)\exp(iH_2t)\exp(-iH_1t)\exp(-iH_2t) \psi_0 \,.
\end{equation}
Expanding in the parameter $t$ (which we may shrink toward zero),
we have
\begin{equation}
\psi = \psi_0 + {t^2 \over 2} [H_1,H_2]  \psi_0 + O(t^3) = 
\exp\left([H_1,H_2] t^2/2 \right) \psi_0 
+ O(t^3) \, .
\end{equation}
If the commutator does not vanish, the alternating application of 
the two operators $H_1$ and $H_2$ at our disposal has
given us a ``new direction'' in which we can move the solution -- 
a new kind of rudder or oar for sailing the seas of Hilbert space.

Going to the case where the number $r$ of controls $i$ is arbitrary (but
finite), it becomes apparent \cite{rama2,lloyd2,rama1,harel,lloyd3,lloyd1} 
that by selecting a sequence of unitary transformations each generated 
by a member of the given set $\left\{ H_i, i=1,\cdots,r\right\}$, one
can achieve the effect of {\it any} Hamiltonian 
in the Lie algebra produced from the original set by an operation
of repeated commutation.  Denoting the skew-Hermitian counterpart
$-iH_i$ of $H_i$ by ${\hat H}_i$, this Lie algebra is the 
real linear vector space spanned by the operators ${{\hat H}_i}$ and
their mutual commutators of all orders.
We can clearly go anywhere in Hilbert space that can be reached through
exponentiation of any member ${\hat L}_i$ of this Lie algebra: 
\begin{equation}
\psi(t) = U(t,0) \psi_0 =  e^{{\hat L}_i t} \psi_0 \, .
\label{eq:exponl}
\end{equation}
This result, rooted in the Baker-Campbell-Hausdorff formula \cite{wilcox}
familiar to physicists, is well known in ``classical'' control 
theory \cite{brockett} and provides the basis for geometric
control.  It is certainly at the heart of the original
effort to extend classical controllability results to the quantum
domain \cite{htc}.  Moreover, it is the key ingredient of Lloyd's
proof \cite{lloyd2} that ``almost all quantum logic gates are
 universal'' (see also Deutsch and coworkers \cite{deutsch1,deutsch2},
Sleator and Weinfurter \cite{sleator}, and DiVicenzo \cite{divicenzo}).

Let us take a moment to understand in broad terms what this last
statement means.  Consider a quantum computer 
that manipulates distinct quantum bits, or ``qubits'' (i.e.,
linear superpositions of definite ``on'' and ``off'' states,
which might be represented by ``up'' and ``down'' states of a 
spin-half particle or the excited and ground states of an
atom).  Such a computer is said to be {\sl universal} if, by implementing
a finite sequence of {\it local operations}, it can perform an arbitrary
unitary transformation over those variables with arbitrary precision.
(This is equivalent to controllability in a finite-dimensional 
state space.)  In particular, one can show that two-qubit CNOT gates, 
combined with single-qubit operations, can do the job.  In this
sense, the CNOT gate is itself said to be universal.  A CNOT gate 
carries out the XOR operation: the XOR of two bits is the sum of 
their Boolean values, modulo 2.  In a fascinating development, 
Gershenfeld and Chuang \cite{gershenfeld} have created a quantum
CNOT gate in the laboratory by exploiting the interaction between
the nuclear spins of the hydrogen and carbon atoms in chloroform
molecules (CHCl$_3$) in a liquid sample subject to an external magnetic 
field.  A sequence of two $\pi$-pulse radiofrequency signals flips the
spin of the H nucleus (representing the target bit) {\it if and only if} 
the spin of the $^{13}$C nucleus (the control bit) is parallel to the 
external field.  This design provides the basis for a quantum computer 
controlled by nuclear magnetic resonance.

By a geometrical argument patterned after that sketched above, 
Lloyd \cite{lloyd2} was able to show that ``almost any quantum logic 
gate with two or more inputs is universal.'' If one can repeatedly 
apply a given control Hamiltonian to a system, which evolves
autonomously between such applications, then provided that the Lie 
algebra generated by the control Hamiltonian and the unperturbed 
Hamiltonian $H_0$ closes on the full space of Hamiltonians for the system, 
one can build any desired unitary transformation on the system.  
``Essentially, any nontrivial interaction between quantum variables 
will do.''  

Against this background, we now resume the main line of development. 
For the control system based on (\ref{eq:nodrift}), strong controllability 
results follow from classical control theory as developed by systems
engineers and mathematicians \cite{chow,sussmann,brockett}.  Introduction 
of an additional term on the right-hand-side of Eq.~(\ref{eq:nodrift}) 
to describe autonomous motion driven by a ``drift'' Hamiltonian $H_0$ 
(which is {\it not} multiplied by a control function)
complicates matters somewhat.  However, this complication has been 
addressed explicitly by Lloyd \cite{lloyd2} and Weaver \cite{weaver} 
and less directly by Ramakrishna and coworkers \cite{rama1,rama2}.

The formulation of Ramakrishna and Rabitz \cite{rama1} is conveniently 
succinct.  They consider a system having an $n$-dimensional state space 
(where $n$ is finite but otherwise arbitrary) and focus on the equation 
of motion of the time-development operator $U$ that ``evolves'' the system
state via $\psi(t) = U(t,0) \psi_0$:
\begin{equation}
{\dot U}(t,0) = {\hat H}_0 U(t,0) + \sum_{i=1}^r u_i(t) 
{\hat H}_i U(t,0) 
\, , \qquad U(0,0) = I \, .
\label{eq:udynam}
\end{equation}
Here $U$ is a unitary $n \times n $ matrix (with $I$ the $n\times n$
identity matrix), while ${\hat H}_0$ and
the ${\hat H}_i$ are $n \times n$ skew-Hermitian matrices.  The
control functions $u_i(t)$ are assumed to be well enough behaved
that the problem (\ref{eq:udynam}) always has a unique solution.
The driftless case where ${\hat H}_0$ is absent, which corresponds 
to Eq.~(\ref{eq:nodrift}), is the one most commonly entertained in 
the theory of quantum computation.  The system (\ref{eq:udynam}) is said
to be {\sl controllable} if the matrices ${\hat H}_0$, ${\hat H}_i$
and the control functions $u_i(t)$ allow every $n \times n$ matrix 
$U$ to be reached in finite time. 
\vspace{2truept}

\noindent
{\sl Theorem \cite{rama1}}.  A necessary and sufficient condition for 
the system (\ref{eq:udynam}) to be controllable is that the set comprised 
of ${\hat H}_0$ and the ${\hat H}_i$ $(i=1,\ldots,r)$, together with 
all commutators and repeated commutators among these matrices (i.e., 
the Lie algebra generated by ${\hat H}_0$ and 
the $\{{\hat H}_i,\,,i=1,\ldots,r\}$) 
equals {\it all} the $n \times n$ skew-Hermitian matrices.  Additionally,
when this condition is met, any unitary $n \times n$ matrix $U$ can be  
constructed by choosing the control functions $u_i(t)$ to be 
piecewise-constant functions of the time. 

\noindent
Further definitive results for the finite-dimensional case have
been derived \cite{schirmer}.

The going gets tougher when the state-space becomes infinite-dimensional
(a true Hilbert space) and especially when one must deal with
unbounded operators and consider the control of dynamical states 
lying in the continuum.  The next section deals with these cases,
which, for brevity, we will refer to as continuous quantum systems,
since they generally involve operators like $x$ and $p$ (position
and momentum) with continuous spectra, as well as eigenstates that
cannot be represented in Hilbert space.  The existing results 
\cite{htc,lloyd3,turinici1,turinici2,turinici3,turinici4} are 
limited, but hardly trivial.

\section{Controllability of Continuous Quantum Systems}

We consider a quantum system whose state $\psi(t)$ evolves 
from $\psi(t=0)=\psi_0$ according to the Schr\"odinger equation
\begin{equation}
{\dot \psi}(t) = \left[ {\hat H}_0 + \sum_{i=1}^r u_i(t) {\hat H}_i \right]
\psi(t)\,,  
\label{eq:basic}
\end{equation}
which differs from Eq.~(\ref{eq:nodrift}) by the insertion of
the usual term for autonomous evolution, or drift.  The state
space ${\cal H}$ can now be infinite-dimensional, and 
${\hat H}_0$, ${\hat H}_1$, ..., ${\hat H}_r$ are linear, time-independent, 
skew-Hermitian operators in this space.  Imposing $||\psi(t)||=1$, the 
system evolves on the unit sphere $S_{\cal H}$ in ${\cal H}$.
As before, the $u_i$ are real functions of $t$.  

Eq.~(\ref{eq:basic}) provides the basis for a rather general control problem.  
In its purest form, the problem is to find a set $u(t)$ of controls $u_i(t)$
that steer the state of the system from $\psi_0$ at the initial time
to a desired target state $\psi_f$ at some later time $t_f$.  One
might alternatively seek controls that lead the state to a 
specified region of the state space, or one might prescribe
a particular trajectory for $\psi(t)$ subsequent to $t=0$.
Thus the quantum control problem is intrinsically nonlinear in that
the controls themselves, which multiply the state $\psi$ in 
Eq.~(\ref{eq:basic}), may depend on the posed behavior of $\psi(t)$.  
However, systems engineers regard this a bilinear control problem: 
bilinear in $\psi$ and the $u_i$.

Transference or extension of results on classical bilinear control 
systems and controllability (see, e.g., Chow \cite{chow}, Sussmann 
and Jurdjevic \cite{sussmann}, and Brockett \cite{brockett}) 
is impeded not only by the infinite dimensionality of the state space 
${\cal H}$ and the unit sphere $S_{\cal H}$, but also by the presence of 
unbounded operators such as $x$, $-i \partial_x $, and $-\partial_x^2$.
These domain problems can be partially overcome if one assumes the 
existence of an {\sl analytic domain} ${\cal D}_\omega$, a set of state
vectors having three properties: (i) it is dense in the Hilbert space 
${\cal H}$, (ii) it is invariant under the given operators ${\hat H}_i$
(with $i=0,1,\ldots,r$), and (iii) on it, the solution of the Schr\"odinger 
equation (\ref{eq:basic}) can be expressed globally in exponential form.  
In simple terms, the availability of an analytic domain, in the context of 
piecewise-constant controls, allows one to write the evolution operator 
$U$ corresponding to a Hamiltonian $H$ in the familiar way, 
$U(t,0)=e^{-iH t}=e^{{\hat H}t}$.  (Strictly, it will also be necessary
to invoke the Nuclear Spectral Theorem \cite{lind} and the construction
of a rigged Hilbert space, to encompass states belonging to continuous
spectra.)

Short of actually finding the controls producing a desired result,
one can ask whether or not such controls exist at all. To address 
this existence issue systematically, we need to adopt precise definitions 
of {\sl reachable sets} and {\sl controllability}.  The state 
$\psi(t)\in S_{\cal H}$ of the controlled system (\ref{eq:basic}) evolves 
from $\psi_o$ on a set ${\rm M}$ which forms a differentiable manifold, 
finite or infinite-dimensional \cite{brockett}.  (We note that 
$S_{\cal H}$ may itself be endowed with manifold structure.) 
\vspace{3 truept}

\noindent
{\sl Def.} Given $\psi_o\,,\, \psi_f \in {\rm M} $, we say that the
state $\psi_f$ is {\sl reachable} from $\psi_o$ at time $t_f>0$ 
if there exists an admissible control $u(t)$ such that
$\psi(t=t_f|u,\psi_o)=\psi_f$.  The set of states reachable from $\psi_o$
at time $t_f$ is denoted $R_{t_f}(\psi_o)$.  The set of states reachable
from $\psi_o$ at some positive time is $R(\psi_o)= \cup_{s>0}R_s(\psi_o)$.
\vspace{3 truept}

\noindent
{\sl Def.} The control system is said to be {\sl strongly completely
controllable} if $R_t(\psi_o) = {\rm M}$ holds for all times $t>0$ and
all $\psi_o\in {\rm M}$.  The system is {\sl completely controllable} if
$R(\psi_o)={\rm M}$ holds for all $\psi_o\in {\rm M}$.

Accommodating the role to be played by the analytic domain, we introduce
modified definitions of controllability:
\vspace{2truept}

\noindent
{\sl Def.}  Let $\psi_o$ be an {\sl analytic vector} belonging to an 
{\sl analytic domain} ${\cal D}_\omega$ that is dense in the state space.  
Then the control system (\ref{eq:basic}) is {\sl strongly analytically 
controllable} [respectively, {\sl analytically controllable}] on 
${\rm M}\subseteq S_{\cal H}$ if $R_t(\psi_o)={\rm M}\cap{\cal D}_\omega$ 
holds for all $t>0$ and all $\psi_o \in {\rm M} \cap{\cal D}_\omega$ 
[respectively, if $R(\psi_o) ={\rm M}\cap{\cal D}_\omega$ holds for all 
$\psi_o \in {\rm M} \cap{\cal D}_\omega$].

The existence of an analytic domain is guaranteed by {\sl Nelson's Theorem} 
\cite{lind}, if we choose to impose the associated conditions, which,
as will now be revealed, are not especially restrictive from the
physical standpoint.  
\vspace{2truept}

\noindent
{\sl Theorem (Nelson).}  Let ${\cal L}$ be a Lie algebra of 
skew-Hermitian operators in a Hilbert space ${\cal H}$, the operator 
basis $\{{\hat H}_{(1)},\cdot\cdot\cdot,{\hat H}_{(d)}\}$, $d< \infty$, 
of ${\cal L}$ having a common invariant dense domain.  If the operator
$T={\hat H}_{(1)}^2+\cdot\cdot\cdot + {\hat H}_{(d)}^2$ is essentially 
self-adjoint, then there exists a unitary group $\Gamma$ on ${\cal H}$ 
with Lie algebra ${\cal L}$.  Let $\bar T$ denote the unique self-adjoint
extension of $T$.  Then it furthermore follows that the analytic
vectors of $\bar T$ (i) are analytic vectors for the whole
lie algebra ${\cal L}$ and (ii) form a set invariant under 
$\Gamma$ and dense in ${\cal H}$.

With the identification
$
{\cal L} = {\cal A} \doteq \{{\hat H}_0, {\hat H}_1, ..., 
{\hat H}_r\}_{\rm LA} \, ,
$
the elements of ${\cal A}$ become densely defined vector fields on 
${\cal D}_\omega \cap {\rm M}$, where ${\dim}\,{\rm M}\cap {\cal D}_\omega 
= d < \infty$ and ${\rm M}$ is the finite-dimensional manifold on which the 
system point evolves with time \cite{brockett}.  The manifold ${\rm M}$ 
is given by the closure of the set 
$ \bigl\{ {\rm e}^{s_0 {\hat H}_{\alpha_0}} 
{\rm e}^{s_1 {\hat H}_{\alpha_1}} \cdot \cdot \cdot 
{\rm e}^{s_r {\hat H}_{\alpha_r}} \psi_o \bigr\}$,
with $(\alpha_0,\alpha_1,\ldots,\alpha_r)$ any permutation of
$(0,1,\ldots,r)$ and $s_k\in {\rm R}^1$, $k=0,\ldots,r$.
Assuming the existence of an analytic domain ${\cal D}_\omega$
(which in general need not entail satisfaction of the requirements 
of Nelson's Theorem), Huang, Tarn, and Clark (HTC) were able to derive 
sufficient conditions for controllability, characterizing the 
reachable sets $R_t(\psi_o)$ and $R(\psi_o)$ in terms of the three 
Lie algebras
\begin{eqnarray}
{\cal A} &=& \{{\hat H}_0, {\hat H}_1, ..., {\hat H}_r\}_{\rm LA}\,, \nonumber \\
{\cal B} &=& \{{\hat H}_1, {\hat H}_2, ..., {\hat H}_r\}_{\rm LA} \, , \nonumber \\
{\cal C} &=& \{{\rm ad}_{{\hat H}_0}^j {\hat H}_i | i=1,...,r; j= 0,1,...\}_{\rm LA} \, ,
\label{eq:liealgebras}
\end{eqnarray}
where ${\rm ad}_X^j Y = [X,{\rm ad}_X^{j-1}Y]$, $j\geq 1$, with 
${\rm ad}_X^0 Y = Y$.  Of special significance are the dimensionalities 
of the tangent subspaces ${\cal A}(\phi)$, ${\cal B}(\phi)$, and 
${\cal C}(\phi)$ of ${\rm M} \cap {\cal D}_\omega$ at 
$\phi \in {\rm M} \cap {\cal D}_\omega$ defined by the vector fields 
associated with these Lie algebras. 

The following key result appears as a corollary of the main theorem
(the so-called HTC theorem \cite{rice3}) proven by Huang, Tarn, and 
Clark \cite{htc}.  The statement and application of the corollary are 
less cumbersome than those of the theorem itself.  
\vspace{2truept}

\noindent
{\sl HTC Corollary 1.}  Let ${\cal C} = \{ {\rm ad}_{{\hat H}_0}^j 
{\hat H}_i | i=1,...,r; j=0,1,...\}_{\rm LA}$ be the ideal in the Lie 
Algebra ${\cal A}= \{{\hat H}_0,{\hat H}_1,...,{\hat H}_r\}_{\rm LA}$ 
generated by ${\hat H}_1,...,{\hat H}_r$.  The system (\ref{eq:basic}),
with piecewise-constant controls, is {\sl strongly analytically 
controllable} on ${\rm M}$ provided that 
(i) $[{\cal C}, {\cal B}] \subset {\cal B}$ 
and (ii) ${\rm dim}\, {\cal C} (\phi) = d < \infty$ for all $\phi \in
{\rm M} \cap {\cal D}_\omega$.

Spelled out, condition (i) of this corollary means that 
$X\in {\cal C}$, $Y \in {\cal B}$ implies $[X,Y] \in {\cal B}$; in 
other words, the Lie algebra ${\cal B}$ must be an ideal in ${\cal C}$. 
Condition (ii) requires that the tangent space associated with 
${\cal C}$ at $\phi$ have constant, finite dimension $d$ for all 
points $\phi$ on the intersection of ${\cal D}_\omega$ and ${\rm M}$.  
If these conditions are met, we can always control the system so that 
the state $\psi(t)$, starting at any point 
$\psi_o \in {\rm M} \cap {\cal D}_\omega$, arrives arbitrarily close to 
any desired point in the (finite-dimensional) manifold ${\rm M}$ after 
any desired time interval $t$.  

Considering that it is applicable to quantum systems having a
state space of infinite dimension, this controllability result 
looks rather positive.  However, within the confines of Nelson's
Theorem and this ensuing result, not all we might desire is within
our grasp.  Intuitively, we realize that an infinite sequence of 
switchings among the piecewise-constant controls would be needed 
to reach an arbitrary goal on the unit sphere in Hilbert space
-- a patently unattainable requirement.
This hard fact is formalized in the following statement.
\vspace{2truept}

\noindent
{\sl HTC Corollary 2 (No-Go Theorem).}  Suppose the set 
$\{{\hat H}_0,{\hat H}_1,...,{\hat H}_r\}$ generates a 
$d$-dimensional Lie algebra ${\cal A}$ which admits an analytic 
domain ${\cal D}_\omega$.  Then the quantum system (\ref{eq:basic}) 
is {\it not} analytically controllable on the {\it full} unit sphere
$S_{\cal H}$ if $d$ is finite.

It is illuminating to consider some simple examples that
satisfy the conditions of HTC Corollary 1 and hence manifest
analytic controllability.
\vspace{2truept}

\noindent
{\it Example 1 (Free Particle)}.  In this simplest of examples,
the Hamiltonian is just $H_0 = p^2/2m$.  Going to skew-Hermitian operators, 
we have ${\hat H}_0 = -i p^2/2m$ and take ${\hat H}_1 = -ip $ and 
${\hat H}_2 = -ix $.  Referring to definitions (\ref{eq:liealgebras}), 
we see that ${\cal B}$ is the so-called Heisenberg algebra \cite{hermann},
which is known to be an ideal in the Lie algebra of all observables 
that are at most of degree 2 in $p$ and $x$ and hence is an ideal 
in ${\cal A}$.  Moreover, ${\cal B}$ is also an ideal in ${\cal C}$, 
satisfying the key condition of HTC Corollary 1.  The eigenstates 
of $p^2$ are of course the plane waves when viewed in the position 
representation, while the operator $x$ generates a shift of momentum 
value from $k$ to $k+ \eta$ via the unitary transformation 
$S(\eta) = e^{i\eta x}$ (with $\eta$ a real parameter).  An analytic 
domain can be constructed from superpositions of plane-wave states 
over finite ranges of momenta.
\vspace{2truept}

\noindent
{\it Example 2 (Rigid Rotor)}.  The Hamiltonian without control is
simply $H_0 = {\bf J}^2 /2I$, where $I$ is the moment of inertia
and the components $J_x$, $J_y$, and $J_z$ of the (purely orbital) 
angular momentum $\bf J$ obey the usual commutation relations
\begin{equation}
[J_x,J_y] = i J_z \,, \quad [J_y,J_z] = i J_x \,,
\quad [J_z,J_x] = i J_y \, .
\end{equation}
Taking ${\hat H}_1 = -iJ_x$, ${\hat H}_2 = -iJ_y$, and ${\hat H}_3 = -iJ_z$, 
the Lie algebras ${\cal A}$, ${\cal B}$, and ${\cal C}$ are seen to
coincide, ${\bf J}^2$ being the Casimir operator of the algebra ${\cal A}$.  
So analytic controllability follows (assuming the existence of an 
analytic domain).  The energy levels $E(J,M_J)$ of the system without 
controls are discrete and are $(2J+1)-$fold degenerate in the magnetic
quantum number $M_J$.  Here $J(J+1)$ and $M_J$ are respectively the 
eigenvalues of ${\bf J}^2$ and (say) $J_z$, with $J$ a non-negative
integer and  $M_J=-J,\ldots,+J$.  The eigenfunctions of $H_0$ and
${\hat H}_0$ in the position representation are the spherical 
harmonics $Y_J^{M_J}(\theta,\varphi)$. 

We encounter here a peculiar situation: although analytic 
controllability strictly holds, it is not possible to change 
the angular momentum quantum number $J$ of the system
with the available controls.  The system evolves on a 
$(2J+1)$-dimensional manifold and it is only possible to change 
the value of the magnetic quantum number $M_J$ (via $J_x$ or $J_y$).
This example illustrates a general property of quantum control
systems:  If the autonomous (or ``free'') evolution is driven by
a Casimir invariant (i.e., if $H_0$ commutes with the controls 
$H_i$), the system state will always remain in the subspace of
a particular eigenvalue of $H_0$ -- even if the technical requirements
for controllability (on the manifold ${\rm M}$!) are met.
\vspace{2truept}

\noindent
{\it Example 3 (Harmonic Oscillator)}.  
The HTC theorem embraces a physical example of prime
importance for physical, chemical, and engineering applications:
namely the simple, one-dimensional harmonic oscillator of mass 
$m$ with coupling to independent external classical fields through 
its position and momentum observables.  With $m=1$ (and $\hbar=1$) 
this problem is mapped into control system (\ref{eq:basic}) 
through the identifications 
\begin{equation}
{\hat H}_0 = -iK_3 \,, \qquad {\hat H}_1 = K_+-K_- \,, \qquad {\hat H_2}
= i(K_++K_-) \, ,
\label{eq:hoops}
\end{equation}
where
\begin{equation}
K_{\pm}= \pm 2^{-1/2} ( \partial_x \mp x ) \, ,
\qquad K_3= (- \partial_x^2 + x^2)/2 \, ,
\end{equation}
while the control functions $u_1(t)$ and $u_2(t)$ are interpreted 
as the external classical fields (assumed piecewise constant in $t$).  
Obviously, the operators $K_+$ and $K_-$ create and destroy harmonic 
excitations (phonons).  The Lie bracket among the ${\hat H}_i$ is 
determined via 
\begin{equation}
[K_3,K_{\pm}] = \pm K_{\pm} \,, \qquad 
[K_+,K_-] = - I \, ,
\end{equation}
where $I$ is the identity operator. 

Our visceral expectation is that the dynamical effect of the drift
operator ${\hat H}_0=-iK_3$ can be cancelled by that of some input that 
dominates ${\cal B}$, assuring strong analytic controllability.  
This judgment is reflected in the geometric analysis of the problem:
It is well known that there is a common dense invariant analytic 
domain ${\cal D}_\omega$ for the operators (\ref{eq:hoops});
adopting the position representation, this domain is spanned 
by analytic functions which are just the Hermite polynomials, 
denoted $\eta_n(x)$, $n=0,1,2,...,\infty$.  On the other hand, 
the Lie algebras ${\cal B}$ and ${\cal C}$ are readily seen to
coincide, so that the required property $[{\cal C},{\cal B}] \subset
{\cal B}$ ensues trivially.
From
\begin{eqnarray}
K_+ \eta_n &=& (n+1)^{1/2} \eta_{n+1}\,,\qquad K_- \eta_n = 
n^{1/2} \eta_{n-1} \, , \nonumber \\
K_3 \eta_n &=& (n+ {1\over2}) \eta_n \,, \qquad  I\eta_n = \eta_n \,,
\end{eqnarray}
it follows that ${\rm dim}\,{\cal A}(\phi) = {\rm dim}\,{\cal B}(\phi) 
= {\rm dim}\,{\cal C}(\phi) = d = 3$
for all $\phi\in {\cal D}_\omega$, and indeed
${\rm dim}\,I({\cal C},\xi) = 3$ for all $\xi \in S_{\cal H} \cap 
{\cal D}_\omega$, where $I({\cal C},\xi)$ is the maximal integral
manifold of ${\cal C}$ passing through $\xi$. 

Corollary 1 of the HTC Theorem clearly applies, and we may conclude 
that (i) the reachable set of $\psi_o$ in $S_{\cal H} \cap {\cal D}_\omega$
is given by $I({\cal C}, \psi_o) = I({\cal B}, \psi_o)$ for
$\psi_o= \psi(x;t=0) \in {\rm span}\{\phi_n(x),n=0,1,2,...\}$, and
(ii) with ${\rm M}$ equal to the closure of $I({\cal B},\psi_o)$, the system 
is strongly analytically controllable on ${\rm M}$. 

It must be emphasized that we have appealed to Nelson's Theorem as 
a vehicle for rigorizing the exponential formula (\ref{eq:uexp}) as
a globally valid expression for the time-development operator $U$.  
Acceptance of the conditions of that theorem necessarily restricts the 
manifold ${\rm M}$ to finite dimension.  It is by no means ruled out that 
stronger controllability results can somehow be derived outside the 
framework of Nelson's Theorem.  In that sense our strategy for handling 
the domain problems arising for unbounded operators may be regarded
as ``overkill,'' especially if one believes that a finite-dimensional
description is sufficient for practical purposes.  We hasten to
add that the analysis reviewed above subsumes the case of a 
finite-dimensional state space considered later by other authors.

It is nevertheless apparent that complete controllability will not
always be attainable, whether for mathematical or practical
reasons.  (For example, NP-complete problems may be encountered 
in attempts to construct optimal controls in systems of any 
complexity, even in the finite-level or finite-dimensional context 
\cite{harel2}.)  Therefore it is sensible to develop ``tailored 
controllability concepts'' \cite{turinici3} suitable for
technologically important problems, including control on specified 
subspaces or manifolds and various forms of approximate control.  

We call attention especially to a recent effort \cite{lloyd3} that 
lays a basis for quantum computation over continuous variables.  The 
notion of universal quantum computation over continuous variables is 
certainly precarious: an infinite number of parameters is required 
to specify an arbitrary transformation over even a single continuous 
variable (e.g.\ position $x$ or momentum $p$); one has no assurance 
that an arbitrary unitary transformation can be approximated by any finite 
number of continuous quantum operations; and the susceptibility to noise 
and decoherence is daunting.  Even so, taking a pragmatic approach and 
implicitly assuming that the domain problems raised above can be sorted 
out satisfactorily, Lloyd and Braunstein show that it makes sense to 
define the notion of quantum computation over continuous variables for 
{\it subclasses} of unitary transformations, in particular, those 
that correspond, through exponentiation {\it \`a la} Eq.~(\ref{eq:uexp}), 
to Hamiltonians $H_i$ that are polynomials of the operators 
associated with the selected continuous variables.  A set of
continuous quantum operations is considered universal with respect
to the given subclass of transformations provided that a finite number
of applications of the operations serves to bring one arbitrarily
close to an arbitrary transformation of the subclass.  Appealing
to the same geometric construction elaborated upon in Sec.~2, it
is straightforward to demonstrate that repeated commutations 
among the ``Hamiltonians'' $\pm x$, $\pm p$, $H= (x^2 + p^2)/2$, 
and $\pm S = \pm (xp + px)/2$ provide for the construction of
any Hamiltonian quadratic in $x$ and $p$ (and of no Hamiltonian
of higher order).  (Note that this intermediate result is of
interest in connection with Examples 1 and 3 presented 
above.)   Introduction of at least one ``nonlinear'' operation is needed 
to build higher-order Hamiltonians; Lloyd and Braunstein find that the 
Kerr Hamiltonian $H^2 = (x^2 + p^2)^2$ suffices for this purpose (though 
any higher-order Hamiltonian will work).  Adding this extra ingredient 
to the mix, repeated commutation allows one to create Hamiltonians 
that are arbitrary Hermitian polynomials of any finite orders in $x$ 
and $p$.  It is asserted that the number of operations increases as 
a small polynomial in the order of the polynomial that is to be formed.  
Extension to the case of many variables $\{x_i,p_i\}$ is achieved through 
the inclusion of a set of interaction Hamiltonians 
$\pm B_{ij} = \pm (p_i x_j - x_i p_j )$.  The authors proceed to 
address issues of electro-optical implementation based on beam 
splitters, phase shifters, squeezers, Kerr-effect fibers, and optical 
cavities, as well as the problem of error correction as a remedy for 
noise.  This effort represents seminal formal progress on the interface 
of quantum control and quantum computation.  

\section{Implications for Quantum Many-Body Theory}

For the many-body theorist, the exciting new developments in quantum 
control and quantum computation offer opportunities to contribute
and opportunities to benefit.

\subsection{Quantum Control}
Laser control of matter requires a subtle cooperation between
the light and matter systems to produce the precise interference
of quantum pathways that is essential to achieving the desired
outcome of an experiment or technological application.  Intuition
is likely to be insufficient for the determination of a proper
laser signal.  This realization has led to the introduction
of systematic theoretical tools for the design of control fields.

One needs to distinguish between {\sl open-loop} and {\sl closed-loop} 
control scenarios.  In closed-loop control, information is extracted from
the output in real time to guide the design of the control field;
whereas in open-loop control one merely imposes a pre-determined
field.  The former is evidently problematic in the case of a quantum
system, since collapse of the wave packet ensues if a measurement 
extracts classical information.  For more discussion of this point see
Lloyd \cite{lloyd1}, who envisions an alternative and potentially more
powerful closed-loop scenario ({\sl quantum feedback control}) in 
which quantum rather than classical information is obtained
and coherence is preserved.

We note, incidentally, that the destructive effect of classical 
measurement is actually innocuous in ``closed-loop'' {\sl learning 
control}, proposed by Judson and Rabitz \cite{rabitz3} and recently put
into practice by Bardeen {\it et al.} \cite{wilson} and Assion 
{\it et al.} \cite{assion}.  In this hybrid experimental/theoretical 
scheme, the system to be controlled ``teaches'' a 
computer-programmable laser pulse shaper to find the optimal control 
field.  A rapid sequence of experiments is performed, in each of which 
the response of the sample is probed and fed into a learning algorithm 
that computes incremental corrections to the control field.  Once the optimal 
field is determined, it can be applied to a ``fresh'' system that 
is not subsequently probed.

Theoretical methods for implementing open-loop control include 
path-planning on unitary groups \cite{rama3,rama4,rama5}, optimal 
control \cite{rabitz1,rice2}, and inverse control \cite{rabitz2}.  
The first of these, intended specifically for finite-level systems, 
is a geometric approach in which controls yielding a target 
time-development operator (and hence a target state) are determined 
explicitly by exploiting the algebraic structure of unitary 
groups under a variety of constraints on the controls.  In the
second method, the variational principle is applied to achieve the
``best possible'' outcome subject to obedience of the Schr\"odinger 
equation and to certain practical constraints on the control fields.  
In the third, one seeks a control that produces a prescribed track 
for the expectation value of some system observable; this approach 
has the advantage that one can solve for the requisite control field 
upon invoking the Heisenberg equation of motion for the expectation 
value \cite{ong1}.  All these methods demand, in principle, complete 
knowledge of the autonomous system Hamiltonian and the field couplings 
(e.g. transition dipole moments).  All require accurate solution of the
Schr\"odinger equation in one form or another.  Implementation of the 
optimal control approach requires solution of a two-point boundary 
value problem involving integration of Schr\"odinger equations
forward and backward in time.  Implementation of inverse control 
requires solution of the Schr\"odinger equation with the control 
present, in order to evaluate certain expectation values entering the 
solution for the control field.  To date, these methods have
only been applied to the simplest of model systems, or the simplest
of molecules.  Beyond these cases, the computational dimensions
become formidable, and the full arsenal of microscopic quantum 
many-body theory will be needed to make significant advances.
The involvement of computational many-body theorists would be
crucial to such an effort.  One of the rewards of the inevitable
iteration process of calculation and comparison with experiment 
(notably in the enlarged context of inverse control) would be an 
enhanced understanding of the system Hamiltonian (e.g.\ in the 
form of electronic energy surfaces) as well as the couplings to the 
external fields.

In similar spirit, the ``closed-loop'' learning-control scheme 
exploits the fact that the system ``knows'' its own Hamiltonian;
the system is made to ``teach'' the necessary aspects of the
autonomous Hamiltonian and field couplings to the laser (and
indirectly the experimenter).

\subsection{Quantum Computation}

The dimension of the space of wave functions of a quantum system
grows exponentially with the particle number, presenting an exponential
barrier to solution by classical computers.  Two decades ago
Feynman \cite{feynman} argued that quantum computers, which by definition
can follow many alternative computational paths simultaneously, should 
be able to overcome this impasse.  He further speculated that there may 
exist quantum computers which can serve as universal simulators of 
quantum systems.  It has recently been argued \cite{lloyd4} that Feynman 
was correct on the second as well as the first count, and that quantum 
computation will permit solution of many-body problems hitherto regarded 
as intractable (e.g. macromolecules, heavy nuclei, hadron structure based 
on QCD).  Procedures have been outlined for efficient simulation of the 
time evolution of Fermi systems (which suffer from the notorious sign 
problem when simulated on a classical computer) \cite{lloyd5}, and for 
exponential speed-up in the determination of eigenvalues and eigenvectors 
(relative to classical computation) \cite{lloyd6}.  However, the conditions 
under which the promise of quantum computation can be realized for strongly 
interacting many-body systems remains to be established in practical 
detail \cite{ortiz}.  In particular, one must consider that the algorithm 
offered in Ref.~51 requires an initial guess for the wave function 
whose overlap with the exact ground state 
does not become exponentially small with increasing particle number.  
It should be fruitful to explore this and other issues raised by 
quantum computation, from the perspective of many-body theory.

\section*{Acknowledgments}
This work was supported in part by the U.~S.~National Science Foundation
under Grant No.~PHY-9900713.  This contribution is also a product of 
research performed under the auspices of the Research Year on ``The Sciences
of Complexity: From Mathematics to Technology to a Sustainable
World,'' held during 2000-2001 at the Center for Interdisciplinary 
Research (ZiF), University of Bielefeld.

\end{document}